
\magnification=1200
\normalbaselineskip=24truept
\baselineskip=24truept
\overfullrule=0pt
\hsize 16.0true cm
\vsize 22.0true cm
\nopagenumbers
\def\lsim{\mathrel{\rlap{\lower4pt\hbox{\hskip1pt$\sim$}}
    \raise1pt\hbox{$<$}}}         
\def\gsim{\mathrel{\rlap{\lower4pt\hbox{\hskip1pt$\sim$}}
    \raise1pt\hbox{$>$}}}         

\def\overleftrightarrow#1{\vbox{\ialign{##\crcr
    $\leftrightarrow$\crcr
    \noalign{\kern 1pt\nointerlineskip}
    $\hfil\displaystyle{#1}\hfil$\crcr}}}
\long\def\caption#1#2{{\setbox1=\hbox{#1\quad}\hbox{\copy1%
\vtop{\advance\hsize by -\wd1 \noindent #2}}}}
\centerline{\bf  A QUARK MODEL OF ANTILAMBDA-LAMBDA}
\centerline{\bf PRODUCTION IN $\bar p p$  INTERACTIONS}
\vskip 18pt
\centerline{M.A. ALBERG}
\medskip
\centerline{Department of Physics, Seattle University, Seattle, Washington

98122 and}
\centerline{Department of Physics, University of Washington, Seattle,

Washington 98195}
\bigskip
\centerline{E.M. HENLEY and L. WILETS}
\medskip
\centerline{Department of Physics, University of Washington, Seattle,

Washington 98195}
\bigskip
\centerline{and}
\medskip
\centerline{P.D. KUNZ}
\medskip
\centerline{Nuclear Physics Laboratory, University of Colorado, Boulder,
 Colorado  80309}
\vskip 24pt
\centerline{\bf Abstract}
\vskip 12pt
A quark model which includes both scalar and vector
contributions to the    reaction mechanism is used in a DWBA
calculation of total and differential cross-sections,   polarizations, and spin

correlation coefficients for the reaction $\bar p p \rightarrow \bar \Lambda
\Lambda$ at  laboratory  momenta from threshold to 1.92 GeV/c. The free
parameters of the
        calculation include the scalar and vector strengths, a quark
cluster size    parameter, and six parameters in the unknown $\bar \Lambda
\Lambda$
interaction. Excellent agreement with   experiment is found for  a
constructive interference of scalar and vector terms, and       for a
$\bar \Lambda \Lambda$ potential which differs from that suggested by several
authors on  the         basis of SU(3) arguments. The fit to the data is better
than

that obtained by other  quark models, which use only scalar or
vector annihilation terms. The agreement        with experiment is as
good as that found in meson-exchange models, which use  more
parameters than the present calculation.
\vfill \eject
\magnification=1200
\normalbaselineskip=24truept
\baselineskip=24truept
\centerline{\bf 1. Introduction}

The reaction $\bar p p \rightarrow \bar \Lambda
\Lambda$ is important for several reasons.
Because polarization and spin correlation coefficients can be
measured through the decays of the $\bar \Lambda$ and $\Lambda$, the
experimental
data  may be more sensitive than others to the underlying mechanism. The

reaction can be described in terms of either meson exchange or
quark models, and may provide a test of which picture is more
appropriate at the momenta and distances which correspond to the
experimental measurements. Because the reaction is very sensitive
to initial and final state interactions it can provide information
about the $\bar \Lambda \Lambda$ interaction, for which there are no direct
experimental  measurements. In the quark model it tests our understanding of
$\bar q q$   annihilation and the creation of a heavy $\bar q q$ pair. Finally,
all of these  questions can be examined because of the extensive, precise data

that has been taken on the reaction by the PS185 collaboration$^1$),
including total and differential cross sections, polarizations and
spin correlation coefficients.

Several groups have used meson-exchange models$^{2-7}$) to
obtain reasonable fits to the data. The leading exchanges in these
models, of $K$  and/or $K^*$ mesons, are of  short range, at  distances
for which one might expect quark degrees of freedom to be
important.  Quark models provide a microscopic picture of the
reaction, which tests our understanding of non-perturbative QCD. In
the simplest quark models either a scalar ( $``^3P_0"$ ) or vector ( $``^3S_1"$
)
interaction is assumed to describe $\bar q q$ annihilation and creation, and

several calculations$^{4,8-13}$) have obtained reasonable agreement
with experiment. In some cases these results have been used to
argue that either the scalar or vector interaction provides the
correct description of annihilation. We have proposed$^{14}$) that both

mechanisms should be included, since by analogy with the $N N$
interaction one would expect both vector exchange (of one or more
gluons ) and a scalar representation of both confinement and
multigluon exchange. The possibility of pseudoscalar exchange  has
also been considered$^{15}$), and it was shown that it may be neglected

relative to the expected dominant scalar and vector exchanges.

In this paper we present the results of a DWBA calculation of
the total and differential cross sections, polarizations, and spin correlation

coefficients that have been measured by PS185 from threshold to
1.92 GeV/c. In section 2 we describe the reaction mechanism that
we use and the quark wave functions on which it operates. In section
3 we consider the initial and final state interactions for the $\bar p p$  and

$\bar \Lambda \Lambda$ pairs. Section 4 includes a description of our fits to
the
data (  both global fits and the best fits at each momentum ) and a
comparison of our predictions to preliminary data at momenta not
included in our fits. We also discuss the parameters of the $\bar \Lambda
\Lambda$
interaction that are derived from our fits. In section 5 we compare
our results to those of other theoretical calculations, both meson-exchange and
quark models,
and discuss the implications of partial  wave analyses of the reaction. Our
conclusions are presented in  section 6.
\bigskip
\centerline{\bf 2. Reaction Mechanism}
\medskip
Our reaction mechanism includes both scalar and vector
contributions to the annihilation and creation of antiquark-quark
pairs. The simplest graphs for these terms are shown in Fig. 1. The
$``^3P_0"$ term represents scalar multigluon exchange and/or the
confining scalar force, whereas the $``^3S_1"$ term represents vector
exchange of one or more gluons. Both terms also include $\bar q q$  pairs in

intermediate states. For $\bar \Lambda \Lambda$ production the active quarks
are
a
$\bar u u$   pair which is annihilated and an $\bar s s$  pair which is
created.
The  spectator quark pairs, $\bar u \bar d$  and $ud$, must each be in an I=0
and
S=0  state, so that the spin of the $\bar \Lambda \Lambda$ pair is carried by
the
strange  quarks. Thus a measurement of the spin state of the  $\bar \Lambda
\Lambda$ pair is a  direct determination of the spin state in which the $\bar s
s$
pair is  created. Both scalar and vector terms are spin triplet, so our
quark model predicts a singlet fraction for the  $\bar \Lambda \Lambda$ pair
which is  identically zero, in good agreement with experiment.  A singlet

contribution can arise, e.g. from a pseudoscalar $``^1S_0"$  term, but this
has been shown to be small$^{15}$).

In our model, the operator for vector exchange is
$$I_v = g_v {\sigma'_3} \cdot {\sigma_3 }\eqno(1)$$

\noindent
and that for scalar exchange is
$$I_s = g_s {\sigma'_3} \cdot \left({\nabla_{3'} -
\nabla_{6'} \over 2m_s} \right) \sigma_3 \cdot \left( {\nabla_3 -
\nabla_6 \over 2m} \right) \,, \eqno(2)$$
\noindent
in which we use $m_s$ = 491 MeV/c$^2$ for the strange quark mass and $m$ = 313
MeV/c$^2$ for the
down and up quark masses. Since $m_s \not = m$, the relative coordinate $R'$
between the lambda
and lambdabar is smaller than the relative coordinate $R$ between the proton
and
antiproton
$$R' = {3m \over (2m + m_s)} R$$
It
appears from Eqs. (1) and (2) that we neglect the momentum-dependence of the
``$^3S_1$" and
``$^3P_0$" propagators, and simply take them to be constants.  However, as we
will show later,
we allow $g_v$ and $g_s$ to be momentum-dependent.  Our matrix element for the
reaction is
$$\eqalignno{{\cal M}_{\bar p p \rightarrow \bar \Lambda \Lambda} & =  \langle
X_{\bar \Lambda
\Lambda} (1'2'3'; 4'5'6')\phi(1'2'3')\phi(4'5'6')|(I_v + I_s) \cr &
|\phi(123)\phi(456)X_{\bar N
N} (123;456)\rangle\,,&(3)\cr}$$
in which $X_{\bar \Lambda \Lambda}$ and $X_{\bar N N}$ are
distorted waves in the relative coordinate between the initial and final
particles,
respectively, and $\phi$ is a Gaussian wave function for the internal motion of
the quarks,
e.g.
$$\phi(123) \, \sim \, \prod^3_{i=1} \, \, e^{-3(r_i - R/2)^2/4b^2}$$

\noindent
in which $R/2$ is the proton center-of-mass coordinate, $r_i$ is the coordinate
of the i'th
quark, and $b$ is the rms radius of the quark distribution of the proton.  We
use SU(3)
symmetric wave functions, so for the lambda and lambdabar $b$ is the rms radius
of the quarks
relative to one-third of the sum of their coordinates, but not their
center-of-mass. \vfill \eject
\centerline{\bf 3. Initial and Final State
Interactions} \bigskip \noindent 3.1. THE $\bar p p$  INTERACTION
\medskip

It is not our intent to find an optimal $\bar N N$ potential. This
problem has been investigated by many authors, and we wish to
concentrate primarily on the reaction mechanism and secondarily on
what can be learned about the $\bar \Lambda \Lambda$ interaction. However,
since
the
results are clearly sensitive to the initial state wave functions,
these must predict reasonable agreement with the $\bar N N$ data available
in the momentum region relevant for the $\bar p p \rightarrow \bar \Lambda
\Lambda$ reaction.
In  most of our work we use the $\bar p p$ potential proposed by Kohno and

Weise$^4$)
$$V_{\bar N N} (r) = U_{\bar N N} (r) + i W_{\bar N N} (r)\eqno(4)$$
in which the real term $U_{\bar N N} (r)$ includes central, tensor, spin-orbit
and
spin-spin terms and the imaginary term $W_{\bar N N} (r)$ is a central
potential

which represents annihilation. The long-range part of $U_{\bar N N} (r)$ is

determined by the G-parity transform of Ueda's$^{16}$) one-boson
exchange potential. The meson masses, coupling constants and form
factors for this potential are given in Table 1. For r $<$ 1 fm each term
in the real part of the potential is smoothly extrapolated to the origin by
means of a
Woods-Saxon form. The imaginary potential $W_{\bar N N} (r)$ is given by:

$$W_{\bar N N} (r) = W_{\bar N N}^{(0)} \left\{ 1 + \exp [(r-r_0)/a]
\right\}^{-1} \eqno(5)$$
\noindent
with $r_0$ = 0.55 fm, a = 0.2 fm, and $W_{\bar N N}^{(0)} = -1.2$ GeV. These
parameters
give good fits to total, elastic, annihilation and charge-exchange $\bar p p$

data  for lab momenta up to 2.5 GeV/c. In the momentum region
relevant for $\bar \Lambda \Lambda$ production experiments, 1.4 to 2.0 GeV/c,
the  $\bar
p p$ data consists of total and elastic cross sections, differential
elastic cross sections, and elastic asymmetry measurements. In this
region the predictions of the Kohno-Weise potential are in
reasonable agreement with the data of Kunne et al.$^{17}$), as shown in Fig. 2.
The fit to
the differential cross sections is better than that to the asymmetry.

Haidenbauer$^7$) has shown that the fit can be improved by adding

spin-dependent terms to the annihilation part of the potential. We also found
that we could
obtain better agreement with the $\bar p p$ data by adding a real part and a
spin-orbit term
to the annihilation potential.  We  did not keep such terms because we wanted
to
work with a
minimal  number of free parameters, and as Timmermans et al.$^5$) have shown, a
good fit to the
$\bar \Lambda \Lambda$ reaction data can be obtained without requiring an
excellent fit to the
$\bar N N$ polarization data.  However, the values of the parameters determined
for the $\bar
\Lambda \Lambda$ interaction are dependent on the model used for the $\bar N N$
interaction.
\bigskip
\noindent 3.2. THE $\bar \Lambda \Lambda$ INTERACTION
\medskip
Our  $\bar \Lambda \Lambda$ potential is initially chosen to be that used by
Kohno
and Weise$^4$), although we find it necessary to vary the parameters
of that potential to get good agreement with experiment. The Kohno-Weise $\bar
\Lambda \Lambda$
potential has the form $$V_{\bar \Lambda \Lambda}(r) = U_{\bar \Lambda
\Lambda}(r) + i
W_{\bar \Lambda \Lambda}(r)
\eqno(6)
$$
in which the real term  $U_{\bar \Lambda \Lambda} (r)$ represents isoscalar
meson exchange
and the imaginary term  $W_{\bar \Lambda \Lambda} (r)$ represents annihilation.
The long-
range part of $ U_{\bar \Lambda \Lambda} (r)$ is derived from the isoscalar
exchanges of
the  Nijmegen YN potential$^{18}$), in which SU(3) relations were used to

determine couplings for the pseudoscalar, vector, and scalar nonets.
The meson masses and coupling constants for  $U_{\bar \Lambda \Lambda} (r)$ are
given in
Table 2. The short-range part is determined by means of a smooth
extrapolation to the origin, as in the $\bar N N$ case. The imaginary term

 W$_{\bar \Lambda \Lambda} (r)$ is taken to be a Woods-Saxon form with the same
radius and
diffuseness as the $\bar N N$ absorptive potential:

$$W_{\bar \Lambda \Lambda} (r) = W_{\bar \Lambda \Lambda}^{(0)} \left\{ 1 +
\exp
[(r-r_0)/a]
\right\}^{-1} \eqno(7)$$
\noindent
The strength $W_{\bar \Lambda \Lambda}^{(0)}$ = 700 MeV was chosen to fit the
$\bar \Lambda
\Lambda$ production  cross-section.

\noindent
The Kohno-Weise potential described above is based in part on
SU(3) symmetry arguments together with the use of the G-parity
transformation, both of which may be questioned. In the absence of
direct experimental data on the $\bar \Lambda
\Lambda$ interaction, this potential is a
good starting point for our analysis. But as we describe in the next
section, good fits to the experimental data require changes in the
parameters of the potential, and thereby give us information about
the $\bar \Lambda
\Lambda$ interaction that has not been previously available.
\bigskip
\centerline{\bf 4. Comparison with Experiment}
\medskip
\noindent
4.1. THE BEST GLOBAL FIT TO THE DATA
\medskip
A nine-parameter fit was made to the PS185 data reported at lab momenta of
1436,
1437, 1445,
1477, 1508, 1546, 1642 and 1695  MeV/c. The 356 data points included in this
set
of
measurements  include differential cross sections, polarizations and spin

correlation coefficients. No data points were excluded from the fits.

Minimization programs which included Monte Carlo, simplex, gradient, and
simulated annealing
techniques were used to search  for the best values of $g_v$  (the strength
of the vector term), $g_s$ (the strength of the scalar  term), $r_0$ (a range
parameter in
the quark Gaussian wave function; $r_0 = 2b/\sqrt3$),  and six parameters in
the
$\bar \Lambda
\Lambda$ potential. Three parameters for the real
part of the potential  were varied: $V$ (the strength of the central
plus spin-spin term), $V_T$  (the strength of the tensor term), and $V_{LS}$

(the strength of the spin-orbit term). Three parameters were varied
in the annihilation term: $W_{\bar \Lambda \Lambda}^{(0)},\,\, r_w$ and $a_w$
(the strength,
radius and  diffuseness of the potential).      We considered three possible
cases for the reaction mechanism:  the vector  $``^3S_1"$ term alone, the

scalar $``^3P_0"$ term alone, and a superposition of both terms. The
results of our searches are shown in Table 3.  All of our searches found minima
for
very small values of $a_w$, the diffuseness of the annihilation potential. We
therefore fixed
$a_w$ at the value of 0.05 fm to avoid reflections from a sharp square-well
potential, which
left 8 free parameters to be varied in our searches.  Clearly the
superposition
of both terms
provides the best fit to the data, with a  $\chi^2$ per data point of 3.2. The
comparison of our
calculation, using  these best global fit parameters, with the experimental
data
is  shown in
Figs. 3, 4, 5 and 6.  The total cross sections are shown in Fig.~3.  Excellent
agreement with
the data is obtained. The differential cross sections are shown  in Fig.~4. We
have excellent
agreement with the data at all momenta,  except for being slightly too low in
the forward
direction at 1546  MeV/c. The polarization measurements, shown in Fig. 5,
provide a  more
stringent test of the theory. Again, we have good agreement  with the data. In
Fig.~6 we compare
our  calculations to the spin correlation coefficients that have been
measured at laboratory momenta of 1546, 1642 and 1695 MeV/c. Here we
have reasonable agreement with the data. Note that some of the
experimental points lie outside the range of physically possible
values [-1 to 1].  We did not discard these points, but included them in our
fits as data
points with the errors reported by the experimentalists.
        In Fig. 7 we show a comparison of the best vector alone, scalar
alone, and combined reaction mechanism calculations with the
experimental data at 1642 MeV/c. Here it is seen that both terms
are needed in the reaction mechanism to get a good fit to both the
differential cross section and polarization data. These results are

representative of all the laboratory momenta studied. Vector or
scalar terms alone can fit the cross sections reasonably well, but
only a superposition fits the spin observables as well.
\bigskip
\noindent
4.2. PREDICTIONS FOR MOMENTA NOT INCLUDED IN THE FIT
\medskip

In Fig.~8 we show the predictions of our model for the
preliminary results that have been taken at 1918 MeV/c. These
measurements include differential cross sections, polarizations, and
spin correlation coefficients. The agreement with the differential
cross section is good, except that the calculation predicts somewhat too large
a
cross section
and a secondary maximum at about 90$^\circ$ which does not exist in the data.
Measurements of
the spin correlation coefficient $C_{xx}$ are well reproduced by our
calculation, but the other
spin correlation coefficients and the polarization are not in good agreement,
which suggests a momentum dependence in the parameters of the fit.  This
momentum, 1918 MeV/c,
is well above the range (1436-1695) MeV/c for which the parameters were fit,
and
above the
threshold for $\bar \Lambda \Sigma$ and $\bar \Sigma \Lambda$ production.
\bigskip \noindent 4.3.
PARAMETERS OF THE $\bar \Lambda \Lambda$ INTERACTION DEDUCED FROM THE FIT
\medskip
        The best global fit parameters given in Table III indicate that
the  $\bar \Lambda \Lambda$ potential differs from that expected on the basis
of
SU(3)
arguments. In the real part of the potential, for which the long-range
behavior is determined by one-boson exchange, the central term is
small. The tensor and
spin-orbit terms are much larger than the predictions of the
one-boson exchange model. This may reflect a greater spin-dependence in the
interaction,
which has also been noted by other investigators$^{5-7}$).  The annihilation
term in the  $\bar
\Lambda \Lambda$ potential  is deeper, longer in range, and much less diffuse
than the
corresponding $\bar N N$ annihilation potential. This variation in potential
parameters is
needed to fit the spin observables, as can be seen in  Fig. 9. A calculation
with the
Kohno-Weise  $\bar \Lambda \Lambda$ potential fits the  differential cross
section at 1695
MeV/c, but fails to predict the  experimentally observed change in sign of the
polarization.
\bigskip
\noindent
4.4.  BEST FITS AT EACH MOMENTUM
\medskip
Our fit to the data at each momentum can be improved by
allowing for a momentum dependence in the parameters of our
calculation. This improvement is shown in Figs.~10-13.  At each
momentum we started with the best global fit parameters, and then
allowed MINUIT$^{22}$) to search for a better fit. The momentum dependence
of the parameters is relatively smooth, except for $g_v$ and the lowest
momenta,
as can be seen
in Fig. 14. The  strength of the scalar term in the reaction mechanism is
relatively
constant, whereas the vector term, except for 1546 MeV/c, increases in strength
as
momentum increases. The real part of the  $\bar \Lambda \Lambda$ potential
remains close
to zero, and the depth of the annihilation term stays relatively constant.  The
strengths of the
tensor and spin-orbit terms show a greater momentum dependence. The ranges of
the quark wave
functions and the Woods-Saxon form of the annihilation potential decrease
slowly
with
increasing momentum. \bigskip
\centerline{\bf 5. Comparison to other Theoretical Calculations}
\bigskip
\noindent
5.1 QUARK MODELS
\medskip
Quark model calculations of differential cross sections and polarizations have
been made by
Kohno and Weise$^4$) and Furui and Faessler$^{12}$).  Kohno and Weise used the
vector
``$^3S_1$" model with the initial and final state interactions we have
described
above.  Furui
and Faessler considered separately the vector ``$^3S_1$" and the scalar
``$^3P_0$" models, with
initial and final state interactions taken from the meson exchange calculations
of Tabakin and
Eisenstein$^2$).  They concluded that their ``$^3P_0$" model was in better
agreement with the
data.  The results of both calculations are compared to ours and to the data in
Fig.~15.  Our
calculations are in better agreement with the differential cross sections.  The
Furui and
Faessler calculations and ours have equally good fits to the polarization, but
Weise
predicts a second ``zero-crossing" in the backward direction, which is not
expected.  Thus, both
scalar and vector reaction mechanisms are needed for a good fit to the data, as
we have argued
above.
\bigskip
\noindent
5.2  MESON EXCHANGE MODELS
\medskip
Meson exchange models have been proposed by Tabakin and Eisenstein$^2$),
Niskanen$^3$), Kohno
and Weise$^4$), Timmermans et al.$^5$), LaFrance et al.$^6$) and Haidenbauer et
al.$^7$).
These models differ in the types of K-mesons included in the exchange (K,
K$^*$,
K$^{**}$) and
in their initial and final state interactions.  The earlier
calculations$^{2,3}$) had a limited
amount of experimental data with which to compare their results.  Kohno and
Weise's meson
exchange results were similar to their quark model calculations, which we have
discussed
above.  A comparison of our results with the later meson exchange calculations
is given in
Fig.~16 for laboratory momenta of 1508 MeV/c and 1695 MeV/c.  Our fits are
better than those of
LaFrance et al. and Haidenbauer et al., each of which fails to reproduce the
steep
forward rise in the differential cross section at 1695 MeV/c and predicts
oscillatory behavior
in the polarization at that momentum which is not seen in the experiment.  Our
fit to the
differential cross section and polarization is as good as that of the Nijmegen
group$^5$). The
Nijmegen fit included the momentum range from 1436 to 1546 MeV/c.  After
excluding 7 data points
they used 10 parameters to fit the 157 remaining data points with a
$\chi^2$/data of 1.2.  We
have not excluded any of the 356 data points in the momentum range from 1436 to
1695 MeV/c in our
9 parameter fit which has $\chi^2$/data of 3.2.  These fits are of essentially
the same quality,
so that the hope that a distinction could be made between the success of the
quark and meson
exchange models for this reaction has not yet been realized.  The definitive
test of these
models may well be at higher momenta and in the $\bar \Sigma \Lambda$ + c.c.
and
$\bar \Sigma
\Sigma$ production reactions.
\bigskip
\noindent 5.3 PARTIAL WAVE ANALYSES
\medskip
A partial wave analysis of the $\bar p p \rightarrow \bar \Lambda \Lambda$
reaction near
threshold has been carried out by Tabakin et al.$^{23}$)  They found that for
momenta up
through 1546 MeV/c, a basis set which includes all LS states with J $\leq$ 1,
plus the
$^3$P$_2$ state, is sufficient to explain general properties of the
experimental
data, such as
the forward peaking of the differential cross section and the zero-crossings in
the
polarization data.  They showed that the use of $^3$S$_1$ and $^3$P$_0$ {\it
{partial waves}}
alone would not produce zero-crossings.  However, they explained that this
conclusion is not
relevant for quark-baked models like ours which use ``$^3$S$_1$" and/or
``$^3$P$_0$" for the
$\bar u u \rightarrow \bar s s$ reaction mechanism, but not for partial waves.
Indeed,
here the quarks are constituents of composite objects, through which higher
partial waves
are brought into the reaction.  We have shown above that in our model the
zero-crossings are well
reproduced.  We included contributions up through L = 7 in our calculations.

In their most recent work Haidenbauer et al.$^7$) have suggested that
spin-transfer observables
should be able to discriminate between meson-exchange and quark models.
However, we are not
certain that this conclusion is correct, in part because their analysis is
based
on a
spin-scattering matrix parametrization that is appropriate for composite
objects, not
constituent quarks, and in part because of the sensitivity to final ($\bar
\Lambda \Lambda$)
and initial ($\bar p p$) state interactions.  We propose a comparison between
the best
meson-exchange calculations and our combined scalar plus vector quark mechanism
rather than the
vector model alone that they used.  Work on this problem is now in progress.

\bigskip
\centerline{\bf 6. Conclusion}
\medskip
We have shown that an excellent fit to experimental data for the reaction $\bar
p p \rightarrow
\bar \Lambda \Lambda$ in the laboratory momentum region from 1436 to 1695 MeV/c
can be obtained
with a quark model that includes both scalar and vector terms in the reaction
mechanism.  This
model is sensitive to both initial and final state interactions.  In order to
achieve good
agreement with experiment, a $\bar \Lambda \Lambda$ potential which differs
significantly from
that expected on the basis of SU(3) arguments is required.  The reaction takes
place
at distances for which quark effects are expected to be important, and our fits
are
comparable to the best of the meson exchange calculations.  In order to
distinguish between these models of the reaction, further comparison is
necessary at higher
momenta and for the production of $\bar \Sigma \Lambda, \bar \Lambda \Sigma$
and
$\bar \Sigma
\Sigma$ pairs.
\bigskip
\centerline{\bf Acknowledgments}
\medskip
We wish to thank the members of the PS185 Collaboration for their interest in
our work.  In
particular, we appreciate discussion of their results with P. Barnes, R.
Eisenstein,
H. Fischer, R. von Frankenberg, D. Hertzog, T. Johansson, K. Kilian and W.
Oelert.  We also
acknowledge stimulating discussions on the theoretical aspects of this problem
with P. LaFrance,
F. Tabakin,  R. Timmermans, and W. Weise.  This work is supported in part by
the
U.S.
Department of Energy.

\bigskip
\centerline{\bf References}
\medskip
\item{1)}P.D. Barnes et al., Phys. Lett. {\bf B189} (1987) 249; Phys. Lett.
{\bf
B229} (1989)
432; Nucl. Phys. {\bf A526} (1991) 575; H. Fischer, Ph. D. Thesis, University
of
Freiburg,
Germany (1992); W. Oelert (private communication)
 \medskip \item{2)}F. Tabakin and R.A. Eisenstein,
Phys. Rev. {\bf C31} (1985) 1857 \medskip
\item{3)}J.A. Niskanen, Helsinki preprint HU-TFT-85-28
\medskip
\item{4)}M. Kohno and W. Weise, Phys. Lett. {\bf B179} (1986) 15; Phys. Lett.
{\bf B206} (1988)
584; Nucl. Phys. {\bf A479} (1988) 433c
\medskip
\item{5)}R.G.E. Timmermans, T.A. Rijken and J.J. deSwart, Nucl. Phys. {\bf
A479}
(1988) 383c;
Phys. Rev. {\bf D45} (1992) 2288
\medskip
\item{6)}P. LaFrance, B. Loiseau, and R. Vinh Mau, Phys. Lett. {\bf B214}
(1988)
317; Nucl.
Phys. {\bf A528} (1991) 557.
\medskip
\item{7)}J. Haidenbauer, T. Hippchen, K. Holinde, B. Holzenkamp, V. Mull and J.
Speth, Phys.
Rev. {\bf C45} (1992) 931; J. Haidenbauer, K. Holinde, V. Mull and J. Speth,
Phys. Rev. {\bf
C46} (1992) 2158.
\medskip
\item{8)}C.B. Dover and P.M. Fishbane, Nucl. Phys. {\bf B244} (1984) 349
\medskip
\item{9)}H. Genz and S. Tatur, Phys. Rev. {\bf D30} (1984) 63; G. Brix, H. Genz
and S. Tatur,
Phys. Rev. {\bf D39} (1989) 2054.
\medskip
\item{10)}P. Kroll, B. Quadder and W. Schweiger, Nucl. Phys. {\bf B316} (1989)
373
\medskip
\item{11)}H.R. Rubinstein and H. Snellman, Phys. Lett. {\bf B165} (1985) 187
\medskip
\item{12)}S. Furui and A. Faessler, Nucl. Phys. {\bf A468} (1987) 669
\medskip
\item{13)}M. Burkardt and M. Dillig, Phys. Rev. {\bf C37} (1988) 1362
\medskip
\item{14)}M.A. Alberg, E.M. Henley and L. Wilets, Z. Phys. {\bf 331} (1988)
207;
M.A. Alberg,
E.M. Henley, L. Wilets and P.D. Kunz, Nucl. Phys. {\bf A508} (1990) 323c
\medskip
\item{15)}M.A. Alberg, E.M. Henley and W. Weise, Phys. Lett. {\bf B255} (1991)
498
\medskip
\item{16)}T. Ueda, Prog. Theor. Phys. {\bf 62} (1979) 1670
\medskip
\item{17)}R.A. Kunne et al., Nucl. Phys. {\bf B323} (1989) 1
\medskip
\item{18)}M.M. Nagels, T.A. Rijken and J.J. de Swart, Phys. Rev. {\bf D12}
(1975), {\bf D15}
(1977) 2547, {\bf D20} (1979) 1633 \medskip
\item{19)}B. Jayet et al., Nuovo Cimento {\bf 45A} (1978) 371
\medskip
\item{20)}B.Y. Oh et al., Nucl. Phys. {\bf B51} (1973) 57
\medskip
\item{21)}J. Button et al., Phys. Rev. {\bf 121} (1961) 1788
\medskip
\item{22)}F. James and M. Roos, Comput. Phys. Commun. {\bf 10} (1975) 343
\medskip
\item{23)}F. Tabakin, R.A. Eisenstein and Y. Lu, Phys. Rev. {\bf C44} (1991)
1749
\vfill \eject
\centerline{Table 1}
\bigskip
\noindent
$\bar N N$ meson exchange potential parameters, taken from
Ueda$^{16}$)\goodbreak \bigskip
\hrule
$$\vbox {\tabskip 2em plus 3em minus 1em
\halign to \hsize{\hfil #\hfill && \hfil #\hfil \cr
Meson&$I^GJ^P$&$g^2$&$f/g$&$\mu$(MeV)&$\Lambda$(MeV) \cr
\noalign {\vskip 12pt}
\noalign {\hrule}
\noalign {\vskip 12pt}
$\pi$&1$^-$0$^-$&-14.01&&138.7&2532.4 \cr
$\eta$&0$^+$0$^-$&2.73&&548.7&1184.3\cr
$\rho$&1$^+$1$^-$&0.778&4.76&763.0&1184.3\cr
$\omega$&0$^-$1$^-$&-8.00&0.0&782.8&1184.3\cr
$a_0$&1$^-$0$^+$&-4.09&&1016.0&1184.3\cr
$f_0$&0$^+$0$^+$&4.44&&1070.0&1184.3\cr
$\sigma$&0$^+$0$^+$&1.96&&416.1&1184.3\cr
\noalign {\vskip 12pt}
\noalign {\hrule}
\noalign {\vskip 12pt}
}}$$
\noindent
$\mu$ and $\Lambda$ are the meson mass and vertex form factor parameter,
respectively.
\vfill \eject
\centerline{Table 2}
\bigskip
\noindent
$\bar \Lambda \Lambda $ meson exchange potential parameters, based on the
isoscalar exchanges of the Nijmegen YN potential$^{17}$)\goodbreak
\bigskip
\hrule $$\vbox {\tabskip 2em plus 3em minus 1em
\halign to \hsize{\hfil #\hfill && \hfil #\hfil \cr
Meson&$I^GJ^P$&$g^2$&$f/g$&$\mu$(MeV)\cr
\noalign {\vskip 12pt}
\noalign {\hrule}
\noalign {\vskip 12pt}
$\eta'$&0$^+$0$^-$&21.6&&957.5 \cr
$\eta$&0$^+$0$^-$&2.10&&548.7 \cr
$\phi$&0$^-$1$^-$&-3.84&2.19&1019\cr
$\omega$&0$^-$1$^-$&-7.73&-0.12&783.9\cr
$f_0$&0$^+$0$^+$&14.0&&1043\cr
$\sigma$&0$^+$0$^+$&5.06&&444\cr
\noalign {\vskip 12pt}
\noalign {\hrule}
\noalign {\vskip 12pt}
}}$$
\vfill \eject
\vskip.75truein
\hskip4.5truein{Table 3}
\bigskip
\noindent
Global fit parameters for searches with fixed $a_w$=.05\goodbreak
$$\vbox {\tabskip 2em plus 3em minus 1em
\halign to \hsize{\hfil #\hfill && \hfil #\hfil \cr
&$g_v$&$g_s$&$r_0$(fm)&W(MeV)&V(MeV)&$V_T$(MeV)&$V_{LS}$(Mev)&$r_w$(fm)&$a_w$(fm
   )&$\chi^2$&$\chi^2$/dat\cr
\noalign {\vskip 12pt}
\noalign {\vskip 12pt}
vector alone&1.1&0.0&0.85&-538&218&111&-25&0.68&0.05&2510&7.1\cr
scalar alone&0.0&4.5&0.56&-1334&-95&38&-69&0.65&0.05&2266&6.4\cr
vector + scalar&-17.4&6.5&0.43&-1956&147&-151&-315&0.66&0.05&1148&3.2\cr
\noalign {\vskip 12pt}
\noalign {\vskip 12pt}
Kohno and Weise Potential Parameters&&&0.64&-700&-268&27&-37&0.55&0.20\cr
\noalign {\vskip 12pt}
\noalign {\vskip 12pt}
}}$$
\noindent
$\chi^2/$dat is the $\chi^2$ per data point of the fit
\vfill \eject

\centerline{\bf Figure Captions}
\medskip
\item{Fig.~1}Lowest order diagrams for $\bar p p \rightarrow \bar \Lambda
\Lambda$
\medskip
\item{Fig.~2}Differential cross section and asymmetry for $\bar p$ laboratory
momentum of 1501
MeV/c.  The experimental data is from Kunne et al.$^{17}$), the calculation
uses
the
$\bar p p$ potential of Kohno and Weise$^4$)
\medskip
\item{Fig.~3}Total cross section for $\bar p p \rightarrow \bar \Lambda
\Lambda$.  The
experimental data is taken from references 1 and 19-21. The theoretical curve
is
calculated
with our best global fit parameters.  The solid part of the curve represents
the
energy region
included in our fit; the dashed part is the extension of the calculation to
energies above that
region.
\medskip
\item{Fig.~4}Differential cross sections for our best global fit parameters.
The experimental
data is from PS185.
\medskip
\item{Fig.~5}Polarization for our best global fit parameters.  The experimental
data is from
PS185.
\medskip
\item{Fig.~6}Spin correlation coefficients for our best global fit parameters.
The
experimental data is from PS185.
\medskip
\item{Fig.~7}A comparison of our best fits to the differential cross section
and
polarization
data at 1642 MeV/c using the vector (dashed line), scalar (dash-dot line) or
combined vector
and scalar (solid line) reaction mechanisms.
\medskip
\item{Fig.~8}The predications of our best global fit for the differential cross
section,
polarization, and spin correlation coefficient data at 1918 MeV/c.
\medskip
\item{Fig.~9}A comparison of calculations done with our best global fit
parameters (solid
line) to a calculation done with the SU(3)-based $\bar \Lambda \Lambda$
potential of
Kohno and Weise$^4$) (dashed line).
\medskip
\item{Fig.~10}Differential cross sections for the best fit parameters at each
momentum.  The
momentum dependence of the parameters is shown in Fig.~14.
\medskip
\item{Fig.~11}Polarization for differential cross sections for the best fit
parameters at each momentum.  The
momentum dependence of the parameters is shown in Fig.~14.
\medskip
\item{Fig.~12}Spin correlation coefficients for differential cross sections for
the best fit parameters at each momentum.  The
momentum dependence of the parameters is shown in Fig.~14.
\medskip
\item{Fig.~13}Our best fit to the experimental data at 1918 MeV/c.
\medskip
\item{Fig.~14}Momentum dependence of the best fit parameters:  14a), the
strengths of the
scalar and vector mechanisms; 14b), the ratios of the strengths of the real,
imaginary, tensor
and spin-orbit terms in the $\bar \Lambda \Lambda$ potential to the values used
by Kohno
and Weise$^4$); 14c), the range of the annihilation potential and the size
parameter of the
baryon quark distributions. \medskip
\item{Fig.~15}Comparison of our calculations (solid line) with other
quark-based
models:  the
vector ``$^3$S$_1$" model of Kohno and Weise$^4$) (solid line) and the scalar
``$^3$P$_0$" model
of Furui and Faessler$^{12}$) (dashed line).
\medskip
\item{Fig.~16}Comparison of our global best fit calculation (solid line) with
the meson-exchange
models of Timmermans et al.$^5$) (long-dash line), LaFrance et al.$^6$)
(short-dash line), and
Haidenbauer et al.$^7$) (dot-dash line).
\bye